\documentclass{article}
\usepackage{emulateapj,epsf,pstricks,apjfonts}

\newcommand{\Msun}{\mbox{$M_{\odot}\;$}}
\def\lsim{\;\raise0.3ex\hbox{$<$\kern-0.75em\raise-1.1ex\hbox{$\sim$}}\;}
\def\gsim{\;\raise0.3ex\hbox{$>$\kern-0.75em\raise-1.1ex\hbox{$\sim$}}\;}
\def \gr{$\gamma$-ray }

\def\beq{\begin{equation}}
\def\enq{\end{equation}}
\def\begar{\begin{eqnarray}}
\def\endar{\end{eqnarray}}
\def\mathnew{\mathsurround=0pt}
\def\simov#1#2{\lower .5pt\vbox{\baselineskip0pt \lineskip-.5pt
        \ialign{$\mathnew#1\hfil##\hfil$\crcr#2\crcr\sim\crcr}}}

\def\cmc{\rm ~cm^{-3}}

\def\kms{\rm ~km~s^{-1}}

\def\etal{{ et al. }}
\slugcomment{Astrophysical Journal, v.538 (in press)}
\begin{document}
\title{\bf 
NONTHERMAL  EMISSION FROM A
SUPERNOVA REMNANT IN A MOLECULAR CLOUD}


\author{A. M. Bykov\altaffilmark{1}, R. A. Chevalier\altaffilmark{2},
D. C. Ellison\altaffilmark{3}, and Yu. A. Uvarov\altaffilmark{1}}
\altaffiltext{1}{A.F. Ioffe Institute for Physics and Technology,  
St. Petersburg, Russia, 194021; byk@astro.ioffe.rssi.ru}
 
\altaffiltext{2}{Department of Astronomy,  University of Virginia, P.O. Box 3818,
Charlottesville,  VA 22903;
 rac5x@virginia.edu}

\altaffiltext{3}{Physics Department, North Carolina State University, Raleigh,
NC 27695; 
 don\_ellison@ncsu.edu}

\begin{abstract}

In evolved supernova remnants (SNRs) interacting with  molecular clouds,
a  highly inhomogeneous structure
consisting of a forward shock of moderate Mach number, a cooling
layer, a dense radiative shell and an interior region filled with
hot tenuous plasma is expected.
 We present a model of nonthermal electron
injection, acceleration and propagation in that environment and find
that these SNRs are efficient electron accelerators
and sources of hard X- and \gr  emission.
A forward shock of velocity $v_s \gsim$ 100 km s$^{-1}$
with an ionized precursor propagating into the
molecular cloud accompanied by magnetohydrodynamic turbulence provides
a spatially inhomogeneous distribution of nonthermal electrons.
The energy spectrum of the nonthermal electrons is shaped by the joint action
of first and second order Fermi acceleration
in a turbulent plasma with substantial Coulomb losses.
Bremsstrahlung, synchrotron, and inverse Compton radiation of the
nonthermal electrons produce multiwavelength
photon spectra in quantitative agreement with the radio and the hard emission observed by
{\it ASCA} and {\it EGRET} from 
IC~443.
We distinguish   interclump shock wave  emission from molecular clump
shock wave emission;  particles reach higher energies in the interclump
shock and that is the likely source of $\gamma$-ray emission and
radio synchrotron emission.
Spatially resolved X- and \gr  spectra from
the supernova remnants IC~443, W44, and 3C391 as might be observed with
{\it BeppoSAX, Chandra XRO, XMM, INTEGRAL} and {\it GLAST} would
distinguish  the contribution of the energetic
lepton component to the $\gamma$-rays observed by {\it EGRET}, constraining
the cosmic ray nuclear component spectra in these SNRs.
These data would provide a  valuable tool for studying the  complex structure 
of molecular clouds where 
SNR radiative shocks interact with dense molecular clumps.

\end{abstract}
 
\keywords{acceleration of particles --- cosmic rays --- ISM: Supernova Remnants
--- radiation mechanisms: nonthermal}

\section{\bf INTRODUCTION}

Supernovae and their remnants have long been identified as
likely sites of galactic cosmic ray acceleration.
Evidence for electron acceleration first came from observations
of radio synchrotron radiation.
The observation of nonthermal X-ray emission from SN 1006 by
{\it ASCA} has provided  evidence for electron acceleration up
to $\sim 100$ TeV (Koyama \etal 1995; Reynolds 1998).
The proton cosmic ray component is more difficult to observe, 
but it has long been recognized that pion decays from collisions
with interstellar gas could give an observable flux of
$0.1-1$ GeV photons.
Initial estimates of $\gamma$-ray emission from supernova remnants (SNRs),
concentrating on the component resulting from pion decays, were made by
Chevalier (1977), Blandford \& Cowie (1982), and
Drury, Aharonian, \& V\"olk (1994).
When $\gamma$-ray emission was apparently detected from SNRs
by {\it CGRO}, the $\gamma$-ray spectrum could not
be fitted by a pure pion decay spectrum and some other component
was needed (Esposito \etal 1996; Sturner \etal 1997;
Gaisser, Protheroe, \& Stanev 1998).
In addition to the pion decays, the relevant processes are
bremsstrahlung emission of relativistic electrons and inverse
Compton emission.
Gaisser \etal (1998) modeled these processes in
detail in order to fit the observed $\gamma$-ray spectra of the remnants
IC~443 and $\gamma$ Cygni.
They assumed acceleration to a power law spectrum in the shock front
and determined the spectral index, electron to proton ratio, and
the upper energy cutoff.

The evolution of the relativistic component in a supernova remnant
has been modeled by a number of groups.
Sturner \etal (1997) assumed particle acceleration in the 
shock front with an $E^{-2}$
energy spectrum and followed the evolution of the particle spectrum.
Their SNR model was homogeneous: a factor 4 density jump at
the shock front, constant density in the interior, and a constant
magnetic field in the interior.
Their shock acceleration model had a diffusion length parameter 
that was important for the
maximum energy reached by the particles.
They modeled the nonthermal emission from IC~443 as synchrotron emission
in the radio and bremsstrahlung in $\gamma$-rays.
de Jager \& Mastichiadis (1997) dealt with the same processes in W44,
as well as inverse Compton emission.
They noted that the observed radio spectrum is flatter than would be
expected from shock acceleration of newly injected particles and
suggested that the particles originated from a pulsar in the supernova
remnant.
Recently, Ostrowski (1999) showed that second-order
electron acceleration by the turbulent medium just after the shock
could flatten the spectra to account for the radio observations of IC~443.
Baring \etal (1999) presented calculations of the broad-band emission
from nonlinear shock models of shell-type SNRs. They used
Sedov adiabatic shock dynamics in a homogeneous medium
and a Monte Carlo simulation
of the particle acceleration taking into account the nonlinear shock structure.
The set of models considered by Baring \etal (1999) covers
the range of  shock speeds 490 $\leq v_S \leq$ 4000 km s$^{-1}$ and
 ambient medium number densities 10$^{-3}\leq n \leq$ 1 cm$^{-3}$.

Massive stars that are the likely progenitors of core collapse supernovae
are expected to be spatially correlated with molecular clouds.
The remnants that are likely to
be $\gamma$-ray sources in {\it CGRO} observations (Esposito \etal  1996)
also show evidence for interaction with molecular gas.
Chevalier (1999) recently studied the evolution of supernova remnants
in molecular
clouds and concluded that many aspects of the
multiwavelength observations could be
understood in a model where the remnants evolve in the interclump
medium of a molecular cloud, which has density of $5-25$ H atoms cm$^{-3}$,
and become radiative at radii $\sim 6$ pc.
The compression in the radiative shell is limited by the magnetic field.
Molecular emission occurs when the radiative shell collides with
molecular clumps.
IC~443 is the remnant with the best evidence for high
energy emission and also shows clear evidence for interaction of the shock
with a molecular cloud  (e.g.,
Burton \etal 1990; van Dishoeck \etal 1993; Cesarsky \etal 1999).

Most of the current models of particle acceleration
in SNRs have dealt with  adiabatic remnants in a homogeneous medium
(see, however,
the model of Boulares \& Cox 1988 for the Cygnus Loop and
the model of Jones \& Kang 1993).
In contrast, our aims here are to examine in detail
the nonthermal emission
of  radiative supernova remnants in molecular clouds using a
kinetic model of electron injection, acceleration and propagation,
including the inhomogeneous structure
deduced from multiwavelength SNR observations.
We use global MHD (magnetohydrodynamic) models of supernova remnants
interacting
with molecular clouds as described by Chevalier (1999), including
the theory of radiative shock structure by Shull \& McKee (1979)
and the kinetic model of electron injection and
acceleration by shocks from Bykov \& Uvarov (1999).
The model described here is relevant to the 
mixed-morphology SNRs (Rho \& Petre 1998), if they are interacting
with molecular clouds.
These remnants comprise a 
sizable fraction of the Galactic SNR population. 

The plan of our paper is as follows.
In \S~2, we treat the energy spectrum of energetic electrons as they
pass through the regions associated with a radiative shock wave.
We include the shock acceleration of particles from the thermal pool.
In \S~3, we present calculations of the nonthermal emission from the
relativistic particles, including emission from shocks in clumps as well
as emission from the radiative shock wave in the interclump region.
Our model is specifically applied to the well-observed remnant IC 443, although
we expect it to more generally apply to other supernova
remnants in molecular clouds.
We also examine the ionization by energetic particles and the energy
involved in the relativistic particles.
A  discussion of results and future propects is in \S~4. 
A detailed 
discussion of expected range of densities, shock velocities and 
magnetic fields determined from multiwavelength observations of 
IC 443, W44 and 3C391 was given in the paper by Chevalier (1999, 
and references therein). 

\section{\bf EVOLUTION OF NONTHERMAL ELECTRONS}

The structure of the flow expected in a radiative shock can be
divided into the following regions (see, e.g., Draine \& McKee 1993):\\
{\it (I)} preshock gas ionized and heated by {\it UV}
(ultraviolet) emission and fast particles from the shock;\\
{\it (II)} shock transition region;\\
{\it (III)} postshock cooling layer;\\
{\it (IV)} cold shell of swept-up gas;\\
{\it (V)} hot low density gas interior to the shock.\\
Radiative shocks are  subject to thermal instabilities
(Chevalier \& Imamura 1982; Bertschinger 1986) as well as
 dynamical instabilities (Vishniac 1983; Blondin et al. 1998).
Although the
effect of inhomogeneous magnetic fields on the stability of radiative shock
flow is yet to be studied, 
MHD turbulence is expected in such a system.
The transport and acceleration of nonthermal particles in the
violent SNR environment are governed by the MHD flow and
depend on the ionization structure and on the  spectrum of   MHD turbulence.
Shocks with velocity $v_{S7} \geq$ 1.1 generate sufficient 
ultraviolet {\it UV} radiation
for full 
pre-ionization 
of hydrogen and helium (He$^+$) in the preshock
region, where the gas temperature is typically 
$\sim 10^4$ K (Shull \& McKee 1979). Here, $v_{S7}$ is the shock velocity
measured in units of $100 \kms$,
and $n_1$ is the ion number density measured in units of 10 cm$^{-3}$.
Shocks with $v_{S7} \sim$ 1.5
propagating in a medium with $n_1\sim$ 1
photoionize gas ahead of the viscous jump
to a depth $N_i~ \gsim 3\times 10^{17}$ cm$^{-2}$.
The shock transition in such a case is collisionless
and supercritical unless the preshock magnetic field
is above $\sim$ 4$\times$10$^{-5} n_1^{0.5}$ G.

The geometry of the magnetic field is important for the
efficiency of  electron injection and acceleration.
It has been argued (e.g., Levinson 1996; McClements et al. 1997)
that quasi-perpendicular shocks are an efficient source of
freshly injected electrons if 
their speed
$v_s \geq \left(m_e/m_p \right)^{0.5} c$;
 the injection efficiency drops sharply below that threshold.
On the other hand, Bykov \& Uvarov (1999) have shown that
quasi-parallel MHD shocks of moderate Alfven Mach number
$M_a \leq \left(m_e/m_p \right)^{-0.5}$ are suitable sites
of  electron injection and acceleration. 
The SNR shock waves in molecular clouds are expected to be in this
regime of Mach number.  Thus we
shall discuss below the quasi-parallel portion
of the shock (magnetic field inclinations $\leq \pi/4$)
propagating into a molecular cloud with $v_{S7} \sim 1.5$.

Nonresonant interactions of the electrons with fluctuations,
generated by kinetic instabilities of the ions in the transition region
of a quasi-parallel supercritical shock, play the main role
in heating and preacceleration of the electrons.
The electron heating and preacceleration (injection)
occur in the collisionless shock front transition region
on a scale $\Delta$ of several hundred inertial lengths
of the ions, $l_i$ (Bykov \& Uvarov 1999).
Here $l_i = c/\omega_{pi} \sim 9\times 10^6~n_1^{-0.5}$ cm
and $\omega_{pi}$ is the ion plasma frequency.
For a SNR shock in a molecular cloud, the scale $\Delta \sim$ 10$^9$ cm
is much shorter then the scales of all the other regions.

To accelerate electrons injected in the shock transition region
to relativistic energy $E(p)$, MHD turbulence should
fill the acceleration region, which has a scale $l_a \sim k_m(p)/v_s$.
Here, $k_m = \max\{k_i \}$, where
 $k_i(p)$ is the diffusion coefficient of an electron of momentum $p$
and $ i= ${\it (I), (III)}
refers to the regions defined above.
To provide the conditions for  efficient
transformation of the MHD flow power to the accelerated electrons,
the MHD turbulent fluctuation spectrum in the vicinity of the shock
should extend to scales
$l\geq r_e(p)$ (resonant scattering), where $r_e(p)$ is the gyroradius of
an electron with momentum $p$.
The MHD fluctuations responsible
for the electron scattering
are collisionless for particles with energies up to a GeV because
the ion-neutral collision length is above 10$^{14}$ cm for
typical parameters of the radiative shock described below (a keV electron has
$r_e \approx $ 10$^8\, [B/10^{-6} \, {\rm G}]^{-1}$ cm).

To calculate the spectra of nonthermal electrons
in the regions {\it i = I $-$ IV},
we use a kinetic equation for
the nearly-isotropic distribution function $N_i(z,p,t)$:
\begin{eqnarray}
   &&     \frac{\partial}{\partial t}\: N_i +
     u_i(z)\: \frac{ \partial}{ \partial z }\: N_i -
     \frac{p}{3} \:\frac{\partial}{\partial p}\: N_i\:
     \left(\frac{ \partial}{ \partial z }\: u_i \right) =      \nonumber \\
    &&
           k_i(p) \: \frac{ \partial^2 N_i(z,p) }{ \partial z^2 } +
      \frac{1}{p^2} \: \frac{\partial}{\partial p} \: p^2 D_i(p) \:
      \frac{\partial N_i}{\partial p} +
      \frac{1}{p^2} \: \frac{\partial}{\partial p} \: [p^2 L_i(p) \:N_i].
\end{eqnarray}
The Fokker-Planck type equation (1) takes into account diffusion and
advection (with bulk velocity $u_i[z]$) of an electron in 
phase space
due to particle interaction with MHD waves and the large scale MHD flow
(Bykov \& Toptygin 1993).
Here, $L_i(p)$ is the momentum loss rate of an electron
due to Coulomb collisions in a partially ionized plasma (e.g., Ginzburg 1979).
The momentum diffusion coefficient $D(p)$ is responsible for 
second order
Fermi acceleration,
and $k_i(p)$ is the fast particle spatial diffusion coefficient.
For low energy electrons 
(i.e., $E\leq E_C$), 
Coulomb and ionization losses
are important in the regions {\it (I $-$ V)}, except for the narrow shock
transition region {\it (II)} where 
acceleration is fast enough to overcome losses
and nonthermal electron injection occurs.
The characteristic energy $E_C$ depends on
the plasma density and ionization state, magnetic field, and fast particle
diffusion coefficients $k_i(p)$ and $D_I(p)$.
The corresponding momentum $p_{ci}$
can be estimated as the point where the last two terms in equation (1) are
equal. We used the energy $E_C$ calculated simultaneously with the
electron distribution function
as a convenient parameter to distinguish between
different possible models of MHD turbulence in the postshock 
cooling layer of the
radiative shock structure (see below in this section).

The diffusion coefficients $k_I(p)$ and $D_I(p)$  depend on
the spectrum of the collisionless MHD turbulence, which is poorly
understood.
A plausible approximation for the diffusion coefficients $k_i(p)$
was assumed here.
For all of the regions we used the
following parameterization:
\begin{equation}
     k_i(p)= k_{i0}  \left\{
   \begin{array}{lll}
         1    ,            &                p_T \leq p \leq p_{\ast} \\
        vp^{a}/v_{\ast}p_{\ast}^a,  &  p_{\ast} \leq p \leq p_{\ast \ast}\\
        cp^2/(p_{\ast \ast}^{2 -a} p_{\ast}^a v_{\ast}) &  p_{\ast \ast} 
        \leq p \leq p_{m}.
   \end{array}
     \right.
\end{equation}
Here, $p_T$ is the momentum of the upstream thermal electrons, 
$p_m$ is the momentum corresponding to 
the
upper cut-off energy $E_m$ 
(see  equation [3]),
and $p_{\ast}$ and $p_{\ast \ast}$ are defined in the next paragraph.
The standard relation for the momentum diffusion coefficient
$D_i(p)$ = $p^2 w_i^2/9 k_i(p)$  (e.g., Berezinsky \etal 1990) was used.
In the low energy
regime $p \leq p_{\ast}$, the particle transport is dominated by
large scale turbulent advection (Bykov \& Toptygin 1993).
The large scale  turbulence is due to stochastic MHD plasma motions
on scales $\Lambda$ longer than the particle mean free path due
to resonant scatterings. The sources of MHD turbulent motions are
the shock wave instabilities mentioned at the beginning of this section.
The large scale vortex rms-velocity $w_i$ typically
is a fraction of 
bulk speed
$u_i$.
The spatial diffusion is energy independent below $p_{\ast}$
and $k_{i0} \sim w_i  \Lambda$.
For $ p_{\ast} \leq p \leq p_{\ast \ast}$, electrons are scattered by
resonant MHD waves presumably generated by the streaming instability
of shock accelerated particles (e.g., Blandford \& Eichler 1987;
Jones \& Ellison 1991). In this case  $w_i$ is close to
the Alfven velocity $v_a$ and $k_i(p)$ is the electron diffusion
coefficient due to resonant wave-particle interactions.  If the
power spectrum of magnetic field fluctuations of Alfvenic turbulence
for the resonant fluctuations is
approximated as $dB^2_k/dk \sim k^{-\theta}$
then the power-law index $a$ (equation [2]) is related to the
index $\theta$ as $a = 2 - \theta$. An important particular case
corresponding to $a = 1$ can be written as $k_i(p) = (1/3) \eta_i v r_e(p)$.
The parameter $\eta_i$ determines the scattering ``strength''
and strong scattering (Bohm limit) corresponds to $\eta_i \sim$ 1.
The diffusion model with $a = 1$ was successfully used for
modeling the observed anomalous cosmic ray proton fluxes in the
interplanetary medium (Ellison \etal 1999) and in the Monte Carlo 
simulations of nonlinear shocks in the shell type SNRs
(e.g., Baring \etal 1999). The momentum $ p_{\ast}$ in our model
can be estimated from the equation
$(1/3) \eta_i v r_e(p_{\ast}) =  w_i  \Lambda$.

MHD waves in the partially ionized plasma of the radiative shock
flow are subject to ion-neutral damping (e.g., V\"olk \etal 1981).
We define the momentum $p_{\ast\ast}$ from the condition
$r_e(p_{\ast\ast}) = \lambda_{in}$, where $\lambda_{in}$ is determined
from the relation $\lambda_{in} = 2\pi v_a/ \Gamma_{in}$ and
$\Gamma_{in}$ is the Alfven wave damping rate due to ion-neutral
collisions (Kulsrud \& Cesarsky 1971).
For $p \geq p_{\ast \ast}$,  electrons are scattered by nonresonant small
scale waves.

The limits imposed on the diffusive shock acceleration of
particles due to ion-neutral MHD wave damping have been discussed by
Draine \& McKee (1993) and  Drury \etal  (1996).
They considered the spectra of Alfven waves driven by  the
instabilities of shock accelerated ions in a partially
ionized plasma. Drury \etal  (1996) obtained
the upper cut-off energy of a proton
due to ion-neutral wave damping
\begin{equation}
E_m \leq v_{S7}^3~T_4^{-0.4}~n_n^{-1}~n_i^{0.5}~{\cal P}^{CR}_{-1}{\rm~GeV},
\end{equation}
where $n_n$ is the neutral particle
density, $n_i$ is the ion density (both are measured in $\cmc$),
and ${\cal P}^{CR}_{-1}$ is the total particle
pressure normalized to 10\% of the shock ram pressure.
The ionization structure of the preshock region is inhomogeneous
so the ion number density, temperature and other numbers
in equation (3) must be taken at an upstream distance
$\sim k_I(E_m)/v_S$ from the collisionless shock transition.
Application to our model yields  $E_m \sim$ 10 GeV 
if $v_{S7}\sim$ ($1.1 - 1.5$) and 
$k_I(E_m) \sim 10^{24}$ cm$^2$ s$^{-1}$
because of the high ionization in the preshock region {\it (I)} and the
postshock cooling layer {\it (III)} provided by the {\it UV} radiation.
One can justify the same
value of
$E_m$ for the electrons because the synchrotron and inverse Compton losses are
relatively unimportant for $E_m \sim$ 10 GeV.

{\it Region I}. In the preshock region of the shock with $v_{S7} \gsim$ 1.1,
we assume a highly ionized plasma of  temperature $T_4 \sim$ 1
 up to depths $\sim$ a few times 10$^{17}$ cm$^{-2}$ upstream.
For the IC~443 preshock region, we adopted
a density $\sim 25$ cm$^{-3}$ and $v_{S7}\approx$ 1.5 (Fesen \& Kirshner 1980).
The magnetic field $B_{\parallel} \approx 10^{-5}$ G (Chevalier 1999), while
$B_{\perp} \approx 5 \times 10^{-6}$ G (the inclination is assumed
to be $\sim \pi/6$).
The Alfven velocity here is about 5 km s$^{-1}$.
We assume a conservative diffusion model
in the shock upstream region where $k_I(p)$ has $a$ = 1 and the
moderate scattering ``strength''
10 $\leq \eta_I \leq$ 100 (Ellison \etal 1999).

{\it Region II}. In the shock transition region, injection and
heating of the electrons occur due to nonresonant interactions with strong
MHD fluctuations generated by the ions. Following the model developed by
Bykov \& Uvarov (1999) we take
\begin{equation}
D_{II}(p) \approx p^2~{\bar C}~ \left(\frac{\delta B}{B_0}\right)^2%
\left(\frac{v_a}{v}\right)^2 \left(\frac{v}{l_i}\right),
\end{equation}
where ${\bar C}\sim 1$   and $\delta B \gsim B_0$.
The Alfvenic Mach number of the shock, $M_a \gsim$ 10, is moderate for our situation.
Nevertheless, the effect of the accelerated particle pressure
on the shock structure could be important (see, e.g., Jones \& Ellison 1991;
Baring \etal  1999).
An exact treatment of the effect is not feasible in our model
because it would  require a kinetic description of
the ion injection.
We accounted for the effect by correction of
the total shock compression ratio $\delta_t$ and
by introducing an extended modified 
shock structure with a precursor and a subshock 
with compression ratio $\delta_s \lsim \delta_t$ 
(e.g., Jones \& Ellison 1991). 
In our case with  Mach numbers
$M \gsim$ 15, the estimated extra compression was about 10\%
and thus $\delta_t\approx$ 4.4. A subshock compression ratio $\delta_s
\approx$ 3.3 is consistent with the assumption of
substantial heating of the precursor gas due to wave dissipation
(Berezhko \& Ellison 1999).
These numbers are in agreement with those from
a hybrid simulation of nonlinear shock structure (Giacalone \etal  1997).
We found that the resulting high energy emission spectra are
sensitive to the particular choice of  precursor flow structure
 in the MeV regime.
The keV and GeV
 emission is not sensitive to the structure if the
compression ratios $\delta_t$ and $\delta_s$ are fixed.

{\it Region III}. The postshock cooling layer has about 
a column density $\sim 3\times$10$^{17}$ cm$^{-2}$
of highly ionized plasma with an initial density about $\delta_t$ times
that of the preshock one.
The Alfven velocity in the highly ionized portion of the region {\it (III)}
is similar to that at
the preshock region, $\sim$ 5 km s$^{-1}$. Large scale vorticity with
an amplitude $\lsim 20\kms$
on scales $\Lambda \lsim 10^{14}$ cm
may occur here because of  shell instabilities of
the radiative shock with $v_{S7} \sim$ 1.5 (see, e.g., section 5
in the review by Draine \& McKee 1993).
This would dominate the low energy electron propagation, providing
a spatial diffusion coefficient of $\sim 10^{20}$ cm$^2$ s$^{-1}$
for the radiative shock parameters described above.

Coulomb losses are important here for  electrons with energies
below $E_C \approx$ 20 keV if the large scale turbulent velocity has a
substantial longitudinal component with $w_{III} \approx 20 \kms$.  
$E_C \gsim$ 8 MeV is expected in the absence of a longitudinal component 
of large scale turbulence in the postshock cooling layer because
the Coulomb losses are overcome by resonant interaction with Alfven 
waves. Since the Alfven velocity is relatively low  ($\sim 5\kms$), 
the value of $E_C$ is much higher in that case.
An accurate description of MHD turbulence in the postshock cooling layer 
is not available now. Thus we considered both limiting cases described above 
and show the uncertainty introduced by the
lack of data concerning  the large scale turbulence 
properties in  Fig. 1. To avoid  overestimating the nonthermal emission, 
we used a conservative minimum value of $E_C$ = 120 keV for the radiative 
shock structure described above, although one could expect 
$E_C \sim$ 20 keV in the most favorable case.

{\it Region IV}. The density and temperature in the radiative shell depend on the
preshock density, magnetic field and  abundances (Shull \& McKee 1979;
Chevalier 1999). From the model of Chevalier (1999) for the
above
preshock parameters,
we adopted $n\approx$ 275  cm$^{-3}$
($N_H \approx 1.6\times$ 10$^{20}$ cm$^{-2}$)
and $T\sim$ 10$^2$ K in the dense radiative shell.
The mass of the dense radiative shell is $\sim 1000$ $\Msun$
and its magnetic field
is $\sim 6\times$ 10$^{-5}$ G.
The ionization in the dense radiative shell is
supported at the level of about a percent by shock accelerated
particles penetrating into the shell (see section 3.5 for details).
The phase velocity of the high frequency collisionless Alfven waves
which are responsible for electron resonant scattering
is about 50 km s$^{-1}$; it is determined by the number
density of the ions.
The Coulomb loss rate of the
electrons  is then dominated by the neutral component.

We numerically calculated  the electron distribution function
in  regions {\it (I-IV)}.
The results were then used in calculations of the nonthermal emission.

\section{\bf NONTHERMAL EMISSION FROM IC 443}

The SNR IC 443 (G189.1+3.0) is a very good candidate for testing our model
because the interaction of the SNR with a molecular cloud was established
from radio and infrared line observations (e.g., DeNoyer 1979;
Mufson \etal 1986;
Burton \etal 1990; van Dishoeck \etal 1993; Richter \etal 1995;
Claussen \etal 1997).
The high energy fluxes are somewhat better determined than
in other cases.
Although we concentrate on IC 443, 
we  briefly mention the case of W44, where
observations of OH masers and CO clumps again provide convincing
evidence of molecular cloud interaction  (Claussen et al 1997;
Frail \& Mitchell 1998). 

To model the nonthermal emission from the shell of IC 443 we integrated
the local emissivities over
the radiative shock structure (regions {\it I $-$IV}).
As a parcel of gas containing accelerated nonthermal electrons
evolves through the radiative shock structure including the dense shell,
the local emissivity from bremsstrahlung and inverse Compton
was calculated using  standard theory (see Appendix A for details).
The electron - ion bremsstrahlung emission differential cross sections
(Bethe-Heitler formulae)
were taken from Akhiezer \& Berestetsky (1957) taking into account the
Elwert factor, which is important for  modeling the keV  emission.
The electron - electron bremsstrahlung contribution to the \gr
emissivity was calculated using  cross sections derived
by Haug (1975).
For the inverse Compton  emissivity of IC 443,
we used the same description
of the background photon field as  Gaisser \etal  (1998).
The hard emission production cross sections used in our calculations are
similar to those used recently by other authors
(Asvarov \etal 1990; Sturner \etal  1997;
Gaisser \etal  1998; Baring \etal  1999).
We also estimated the contribution of pion decays to the \gr emission
using a model for the proton component and the cross sections
from Dermer (1986).
The resulting 
emission in the {\it EGRET} regime
 was roughly 7 times below  that from electron bremsstrahlung,
 which is consistent with the results of Sturner et al. (1997).
 We do not discuss the pion component further.

\vspace*{-1.2cm}
\centerline{\null}
\vskip3.55truein
\includegraphics{fig1.pstex}
\figcaption{
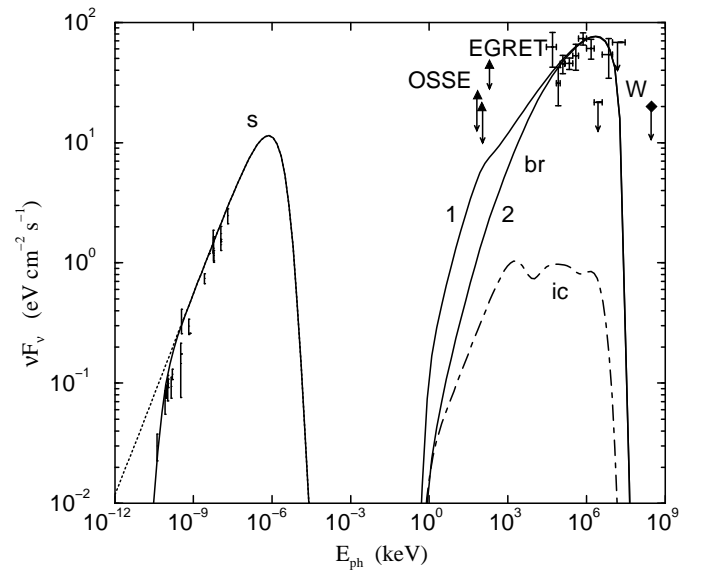
Broadband $\nu F_{\nu}$ spectrum of
the shell of IC 443 (distance 1.5 kpc) calculated from a model of
nonthermal electron production by a radiative shock
with direct injection of electrons from the thermal pool.
The  shock velocity is
150 km s$^{-1}$, the interclump number density is 25 cm$^{-3}$, and
the interclump magnetic field is $ 1.1\times 10^{-5}$ G.
The electron diffusion coefficient (see equation [2]) is
$k_{III0} = 1.1 \times 10^{20}$ cm$^{-2}$ s$^{-1}$, $a=1.0$,
$E(p_{\ast}) = 1$ MeV, and $E(p_{\ast\ast}) = 20$ GeV 
(equation [3]).
The radiative shell parameters are described in
\S~2. The two solid curves labeled by numbers 1 and 2
 correspond to two limiting values of $E_C$: 
$E_C$ = 120 keV (curve 1) and $E_C$ = 2 GeV (curve 2).
Interstellar absorption of the radio spectrum
is treated as described in the text.
The extended inverse Compton emission  from the whole
remnant is shown 
as a dashed line.
The observational data points
are  from Esposito \etal (1996) for the {\it EGRET}
source 2EG J0618 +2234 and Erickson \& Mahoney (1985)
for the IC 443 radio spectrum. Upper limits
for MeV  emission are from {\it OSSE} observations
(Sturner \etal 1997). Upper limits  for $\gamma$-rays $\geq$ 300 GeV  
are from {\it Whipple} observations of IC 443 (Buckley \etal 1998). \label{fig:zeta0} }
\vskip0.2truein

The synchrotron emission of the electrons calculated using the
scheme described by Ginzburg (1979) can be compared
to the radio observations and
to the upper limits at higher frequencies. We also included
 free-free absorption of low frequency radio waves
(dotted line in Fig. 1).
The synchrotron losses of a relativistic electron were included 
following Ginzburg (1979);  bremsstrahlung
energy losses are relatively unimportant in our case.

\subsection{Energetic Nonthermal Emission from the Shell}

The emission was integrated over the regions of the radiative shock
in order to model the spectrum of the entire shell.
Region {\it IV} is the dominant contributor to the synchrotron
and bremsstrahlung emission.
The results in Fig. 1 can be compared to the {\it ASCA} and {\it EGRET}
observations of
SNRs, as well as to the significant upper limits that
have been set at TeV energies by Whipple observations (Buckley \etal  1998)
and from the preliminary analysis of {\it CGRO OSSE}
observations presented by Sturner \etal  (1997).
We tried different parameter sets to fit the observations.

In  Fig. 1, $\nu F_{\nu}$ fits to the observations of IC 443
 are presented for the model
described in the previous section.
The preshock density is taken to be $n = 25 \cmc$
and the assumed distance to IC 443 is 1.5 kpc (e.g., Fesen \& Kirshner 1980).
Interstellar photoelectric absorption of the X-ray spectrum
is accounted for with a line-of-sight
column density $N_{H} = 2\times 10^{21}$ cm$^{-2}$ using the cross sections from 
Morrison \&  McCammon (1983).

For $v_{S7}$ = 1.5 in the diffusion model described by equation (2), we used
 $k_{III0}=  1.1\times 10^{20}$ cm$^2$ s$^{-1}$, $a$ = 1.0,
$E(p_{\ast}$) = 1 MeV, and $E(p_{\ast\ast}$) = 20 GeV,
which are compatible with theoretical estimates of resonant
wave-particle interactions. 
The corresponding value $\eta \approx$ 30.
While in principle $\eta$ can be determined from observations of
heliospheric shocks, single spacecraft observations do not provide
direct values and are subject to ambiguous interpretation. Values of
$\eta \sim 30$ are consistent with modeling of the highly oblique
solar wind termination shock  (Ellison \etal 1999), but smaller values
($\eta \sim 3-10$) have been inferred for interplanetary traveling
shocks (Baring et al. 1997). On
the other hand, very weak scattering ($\eta \sim 100$) has been
inferred from observations near corotating interaction regions (e.g.,
Fisk,  Schwadron,  \& Gloeckler 1997) and
in the solar wind when the interplanetary magnetic field is nearly
radial (e.g., M\"obius et al. 1998). 
We use $\eta \approx$ 30 as a conservative value.
If strong MHD turbulence is present in the upstream shock  region,
the $\eta$ values
could be closer to 1.  This might allow an electron acceleration model
similar to that described earlier, but for slower shock  velocities 
($\lsim 100\kms$).

The diffusion propagation model is not unique.
We obtained a similar fit to that presented in Fig. 1
for a diffusion model with
$k_{III0}$ = 10$^{20}$ cm$^2$ s$^{-1}$, $a$ = 0.7,
$E(p_{\ast}$) = 1 MeV, $E(p_{\ast\ast}$) = 20 GeV. 
As  mentioned in \S 2, there is an important 
uncertainty concerning the lack of a quantitative model for the  generation of
large scale MHD turbulence  in the postshock  cooling 
layer. The uncertainty affects the hard X-ray spectral calculations.
In view of this, we calculated the expected 
bremsstrahlung emission for two limiting cases covering the range 
of uncertainties. Curve 1 in  Fig. 1  corresponds to fully
developed Alfvenic large scale MHD turbulence  in the postshock 
cooling layer ($E_C$ = 120 keV). Curve 2 corresponds to the case of a
lack of large scale MHD turbulence ($E_C$ = 2 GeV).
While the GeV  bremsstrahlung emission (as well as 
synchrotron and inverse Compton emission) is similar in the 2 cases, 
 the hard X-ray emission is  sensitive to the uncertainty.
We also obtained a similar fit to that presented in Fig. 1
for a model with $v_{S7}$ = 1.1 and a diffusion model described by
 $k_{III0} =  5\times 10^{19}$ cm$^2$ s$^{-1}$, $a$ = 1.0,
$E(p_{\ast}$) = 1 MeV, and $E(p_{\ast\ast}$) = 10 GeV. The model
with $v_{S7} \approx$  1.1 requires a somewhat higher
power conversion efficiency than that with $v_{S7} \approx$ 1.5
(see \S 3.6), but this value is preferred for the IC 443
radiative shock velocity (Chevalier 1999).

The model of electron injection from the cloud thermal pool
described above is based on the ionization structure of a
radiative shock, which is very sensitive to the shock velocity
if $v_{S7} \sim$ 1 (Shull \& McKee 1979; Hollenbach \& McKee 1989).
In this respect, it is instructive to consider  a model
in which high energy electrons are not related to freshly injected 
particles, but  are reaccelerated cosmic ray electrons 
(Blandford \& Cowie 1982; Chevalier 1999; Cox \etal 1999).
To model that case we suppose that the
far upstream electron flux is just that observed for Galactic
cosmic ray electrons near  Earth, but extrapolated back to
the energy $E_{crm}$ with the same slope as observed above a GeV.
The calculations of $\nu F_{\nu}$ for this case
  are shown in Fig. 2. We consider the radiative shock
 structure with the parameters described in  \S 2 (the same as
 were used for Fig. 1). The Galactic cosmic ray electron flux in
  the far upstream region was taken to be 3$\times 10^{-2} (E/{\rm GeV})^{-3}$
 e$^{\pm}$ s$^{-1}$ cm$^{-2}$ GeV$^{-1}$ sr$^{-1}$ (e.g., Berezinsky \etal 1990)
 for $E\geq E_{crm}$.  We assumed a flat cosmic ray 
 electron spectrum below $E_{crm}$, 
 corresponding to  flattening due to Coulomb losses in
 the interstellar medium.
 The maximum energy of cosmic ray electrons reaccelerated by the radiative
 shock is $E_m \sim$ 10 GeV (see equation [3]), as in the case shown in Fig. 1.
 The energy $E_{crm}$ was considered as a free parameter here.

\vspace*{-1.2cm}
\centerline{\null}
\vskip3.55truein
\includegraphics{fig2.pstex}
\figcaption{
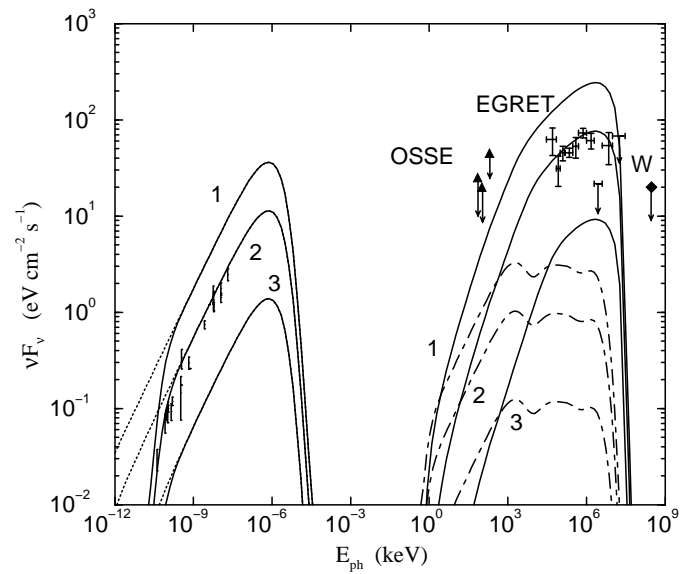
Broadband $\nu F_{\nu}$ spectrum of
the shell of IC 443 calculated for the model of
Galactic cosmic ray electron reacceleration by a radiative shock.
The basic shock parameters are as in Fig. 1.
The low energy cut-off of the preexisting Galactic cosmic ray
electron spectrum in the cloud is  3 MeV for  curve 1,
 8 MeV for  curve 2, and  50 MeV for  curve 3.
The  inverse Compton emission  from the whole
remnant is shown as 
dashed lines.}

\vspace*{0.5cm}

We do not distinguish between electrons and positrons
  because there is no difference between them
 in the MHD shock acceleration process considered above and most of
 the emission from high energy leptons involved in our model
 remains the same for both kind of particles.
 Because of  interactions with the radiative shock structure,
 accelerated cosmic ray electrons could provide a good fit for 
 both synchrotron 
 radio and {\it EGRET}  emission if $E_{crm}\leq$ 10 MeV
 for the radiative shell parameters described above, but fall 
 short for $E_{crm}\approx$ 100 MeV.

 The relatively low
 energies, $E_{crm}\sim$ 10 MeV, required for the IC443 shell
 are a potential problem for this model.
The radio synchrotron radiation from the Galaxy indicates that
the cosmic ray electron spectrum flattens to $N(E)\propto E^{-2}$ below
energies $\sim 1$ GeV (Webber 1983).
If we use such a spectrum in our model, the nonthermal emission
is significantly below that observed.
However, there are uncertainties about betatron-type acceleration
in the shock compression, about the magnetic field strength, and
about the volume of the emitting region.
Duin \& van der Laan (1975) did successfully model the IC 443
radio flux with a cosmic ray compression model.
We found that models with a higher magnetic field and/or larger volume
could reproduce the radio flux, but that the $\gamma$-ray fluxes observed
by {\it EGRET} could not be achieved.
 Cox \etal (1999) have recently concluded
  that cosmic ray compression by radiative shocks
 could provide a good fit to radio and gamma-ray emission from W44
 if the cosmic ray electron population in the vicinity of the SNR
 is higher than that of measured in solar neighborhood.
They used a somewhat higher magnetic field
 ($B\sim$ 200 $\mu$G) in the radiative shell of W44
 than the 60 $\mu$G field we assumed for the IC 443 shell. 

The acceleration time is a constraining parameter
in this model because of the relatively low shock speed $v_{S7} \sim 1.5$
in the dense medium. The test particle shock acceleration time $t_a(E)$
 can be estimated from the equation
\begin{equation}
t_a = \frac{3}{v_I - v_{II}} \int_{p_0}^p \left( \frac{k_I(p)}{v_I} + \frac{k_{II}(p)}{v_{II}}\right) \frac{dp}{p}
\end{equation}
where the velocities $v_{i}$ are measured in the shock rest frame
(e.g., Axford 1981).
For the radiative shock parameters described above, the acceleration time
for $E \gsim$ 10 GeV is about 3,000 years, for $\eta \sim$ 30 (note that $t_a \propto \eta$),
consistent with
the age of IC 443 estimated from the
hydrodynamical model by Chevalier (1999).
On the basis of {\it Ginga} observations of IC 443, Wang \etal (1992)   suggested
an age of about one thousand years and an association
with the supernova of AD 837.
The reasons for the low age 
are related to a thermal interpretation of the observed 
X-ray emission above 10 keV.
We shall argue below for the nonthermal origin of the hard emission
in a radiative shock model.
A distinctive feature of the radiative shock SNR shell model with direct injection
of the electrons from the thermal pool
is a hard spectrum of
nonthermal emission extending from keV to GeV energies (Fig. 1). 
There is also a hard emission component
due to inverse Compton emission of relativistic electrons having a
scale comparable to the scale of the whole radio image of the remnant.

\subsection{Radio Emission from the Shell}

Observations of the radio emission from IC 443 were performed
with different instruments  over the last
35 years (see, e.g., Erickson \&  Mahoney 1985; Green 1986;
Claussen \etal 1997 and references therein).
The 151 and 1419 MHz radio maps (Green 1986)
reveal that the most intense flux comes from the shell of the remnant.
We suggest that the eastern part of the shell  
is a radiative shock which
has a high magnetic field and electron flux, dominating
the synchrotron radiation from the remnant.

We calculated the synchrotron radiation from the remnant using the local
emissivity given by equation (A17). The integrated spectrum from the remnant
is dominated by the emission from the radiative shell
(having magnetic field $B \sim 6\times 10^{-5}$~G) if the
magnetic field in the central parts of the remnant (region {\it V})
is moderate, $B\le  4\times 10^{-6}$~G.
For modeling  the shell emission, we took into account the interstellar
free-free absorption at low radio frequencies,
assuming a column density to IC 443 of
$N_H\approx 2\times 10^{21}$ cm$^{-2}$
and an average interstellar medium
temperature $T\approx 10^3$ K.
This column density is
similar to that adopted for the recent analysis
of {\it ASCA} data by Keohane \etal (1997) and is about four times
less than the value used earlier
by Petre \etal (1988) and Wang \etal (1992). We assumed that 
the ionization fraction 
is $\sim 15$\%.
The  free-free absorption inside the 
cool radiative shell is not very important above
50 MHz because the emission measure   is about 1 cm$^{-6}$ pc.
The solid and dot-dashed lines
on  Fig. 1 show the total synchrotron radiation from IC 443
(dominated by shell emission in our model)
expected at the Earth compared to the observational data 
(Erickson \& Mahoney 1985). Both of our models are illustrated:
 direct injection of thermal electrons (Fig. 1) and  reacceleration
of preexisting cosmic ray electrons by a radiative shock (Fig. 2).
A radiative shock with total compression ratio $\delta_t$ = 4.4,
including a shock precursor and subshock ($\delta_s$ = 3.3), was considered.
An important feature of the radio observations of IC 443 and W44 is that
their spectral indices are relatively flat, implying  electron
energy spectra flatter than those produced by standard strong shock acceleration in
the test particle limit. In our model, the calculated radio
spectrum  fits that observed from the IC 443 shell 
because of the large shock compression ratio $\delta_t \approx 4.4$
 and to a lesser extent because
of the effect of second order Fermi acceleration
discussed recently by Ostrowski (1999).

\subsection{Nonthermal Emission from a Shocked Clump}

Molecular clouds show complex structure with an ensemble of dense
massive clumps embedded in the interclump medium (Blitz 1993).
We considered  the possibility of  radiative shock interaction
with dense molecular clumps, resulting in high energy emission.
In IC 443 and W44, those regions do not appear to be significant
sites of GHz radio continuum emission,
but the clumps may contain most of the
cloud mass and are  possible sites of
hard X-rays and MeV  $\gamma$-ray emission.
The interaction of a radiative shell with a molecular clump generates 
a dense slab bounded by two shocks (Chevalier 1999). 
The forward 
shock propagating into the clump is strong (see also the simulations
of Jones \& Kang 1993).
Since the clump size is
typically smaller or comparable to that of the radiative shock layer,
it is embedded
in a pool of nonthermal particles created by the radiative shock 
according to the scenario described above. Accelerated particles
as well as ionizing radiation from the
hot interior regions of the SNR should
provide ionization of the external portions of a dense clump
at the level $\gsim 1$\%. This is sufficient for a shock propagating
into the dense clump with $n\sim$ 10$^4$ $\cmc$ and  $B\sim$ 0.1 mG
(e.g., Claussen \etal 1997)
to be collisionless. The percent ionization level provides
 Alfven Mach numbers $M_a \gsim$ 1 and
weak damping of collisionless
Alfven waves up to wavelengths $\sim$ 10$^{10}$ cm for a
flow with velocity $\gsim$ 25 $\kms$ in the clump.
Then the injection and acceleration
processes considered above would be efficient and the ionization in the
preshock region would be maintained self-consistently at least at a percent
level even for a shock propagating inside the dense clump.

The gas ionization by UV radiation in the radiative shock precursor
is sensitive to the shock speed if $v_{S7} \leq$ 1.1 (Shull \& McKee 1979; Hollenbach \&
McKee 1989).  This implies
that a shock having $v_{S7} \leq$ 1 propagating in a dense clump
would be able to accelerate
electrons only below $E_m <$ 1 GeV (see equation [3]). We shall estimate below also
the gas ionization rate due to accelerated nonthermal particles.
The lifetime of a pre-existing
electron of MeV energy against  ionization losses in 
the interclump medium ($n\sim$ 10 cm$^{-3}$) is about $ 10^4$ years.    
Unless a nearby source of fresh accelerated electrons is present, 
Galactic cosmic ray MeV  electrons are  unlikely to be present in the cloud. 
Direct injection
of electrons from the thermal pool of the weakly ionized precursor 
would be a dominant source of shock accelerated electrons up to 
the cut-off energy $E_m$. Because of a lack of accelerated GeV  
electrons in that case ($v_{S7} \leq$ 1) the source would not appear
as a bright radio and hard {\it EGRET}  \gr emission feature.
Nevertheless, these shocks might manifest themselves
as  hard  X-ray and
MeV  \gr emission
 sources 
 with $\nu F_{\nu} \propto \nu^a$ $(a \gsim 0.5)$ 
up to a few MeV  and with a soft spectrum at higher energies.
A comparison of Fig. 3  with Fig. 1 shows that the hard X-ray
flux at 7 keV and above could be dominated by the emission from the
shocked clumps, which is consistent with {\it ASCA} observations of IC 443
presented by Keohane \etal (1997). 
The hard X-ray nonthermal emission flux may be variable on
a timescale of years.
A relatively high MeV emissivity is
a distinctive feature of shocked clumps, but they are difficult to observe.
They must be nearby
(within a few hundreds of parsecs) to be detected with
{\it COMPTEL/OSSE CGRO}.   

For a forward shock of velocity $v_{S7} \leq$ 1 propagating into
a dense clump of number density $n \sim$ 10$^3$ cm$^{-3}$
with upstream ionization level of a few percent (at $\sim k_I[E_m]/v_S$
distance), the upper cut-off energy $E_m \lsim$ 0.5 GeV.
This is about 40 times less than
that for a radiative shock in the interclump medium.
The spatial diffusion $k_{i0}$ below $p_{\ast}$
 scales as $w_i  \Lambda$. Since the scale of a postshock
radiative cooling layer $\propto n^{-1}$ and $\Lambda$ is a fraction
of that, we have $k_{III0} \sim  3 \times$ 10$^{18}$ cm$^2$ s$^{-1}$ in the
clump case.

The effective compression of the nonthermal electrons as well as the magnetic
field by the system of two shocks bounding the slab could be  high.
The magnetic field in the clump is expected
to be high.
Claussen \etal (1997) derived line-of-sight magnetic field strengths
of $\sim$0.2 mG, remarkably uniform on the scale of several parsecs.
Relativistic electrons accelerated by a shock propagating into a clump
should produce radio emission with a flat spectrum up to some hundreds of
MHz, which is lower than the maximum radio
frequency from the shell because of the lower $E_m$ in the clumps.
The emission measure of a clump interacting with a radiative shock
could be EM$ \lsim$ 1,000  cm$^{-6}$ pc, which is
much higher than that of the shell. Thus, the internal free-free absorption
could be important here providing a flatter radio spectrum in the 100s of MHz
regime. This is possibly a reason for the observed spatial variations
of the radio spectral index in IC 443 (e.g., Green 1986).

We present in Fig. 3
the calculated $\nu F_{\nu}$ spectrum of nonthermal
emission from a radiative shock - molecular
clump interaction region. The forward shock velocity  was high -- $100 \kms$,
the initial number density in the clump 1,000 cm$^{-3}$, and
the shocked clump radius 0.5 pc. The shock compression ratio was the same as
that in the radiative shell model:
 $\delta_s \approx$ 3 and $\delta_t \approx$ 4.4.
The ionization of gas in the shock upstream was about 5\%
(at the distances $\sim k_I[E_m]/v_S$). To account for some uncertainty in the 
gas ionization, we present the radio spectra for two possible 
values of the emission measure EM of the shocked clump:  
EM = 110 cm$^{-6}$ pc (solid line) and EM = 900 cm$^{-6}$ pc (dashed line).
The diffusion coefficient was
$k_{III0} \sim  3 \times$ 10$^{18}$ cm$^2$ s$^{-1}$,  $E(p_{\ast}$) = 1 MeV. 
We illustrate the results 
for two possible maximal energies of the electrons 
(depending  on the time of shock - clump interaction). Curve 1 
in  Fig. 3 corresponds to $E_m =$ 0.5 GeV and $E_m =$ 0.05 GeV for
curve 2. 
The synchrotron radio spectrum for  curve 2 
is 
below the scale of the plot. The radio spectrum corresponding to 
curve 1 
is very hard because of internal absorption in the shocked clump medium.

\vspace*{-1.2cm}
\centerline{\null}
\vskip3.55truein
\includegraphics{fig3.pstex}
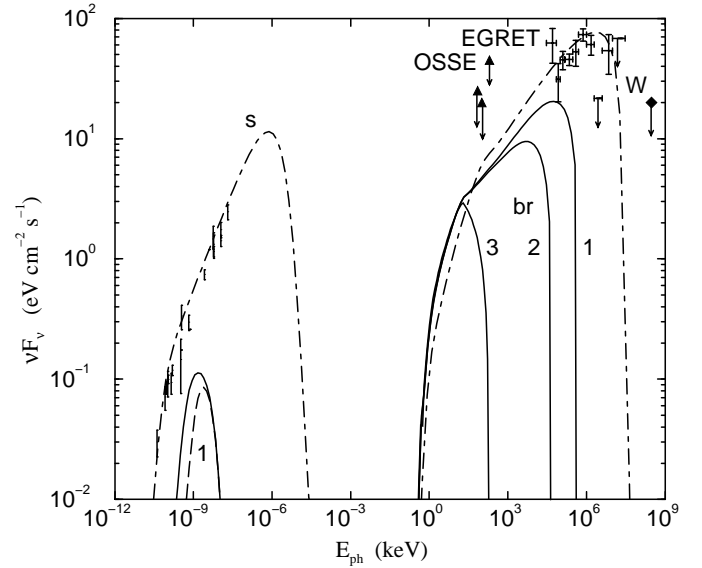
\figcaption{Broadband $\nu F_{\nu}$ spectrum  for a model of
nonthermal electron production by a radiative shock
interacting with a molecular clump (distance 1.5 kpc).
The model spectra are
calculated for the case of a shock velocity in the clump of
100 km s$^{-1}$, a number density of 10$^{3}$ cm$^{-3}$,
and a magnetic field strength of $2\times 10^{-4}$ G.
The electron diffusion coefficient (see equation [2])
$k_{III0} = 3 \times 10^{18}$ cm$^{-2}$ s$^{-1}$, $a=1.0$, and
$E(p_{\ast}) = 1$ MeV. The maximum energy of
accelerated electrons is $E_m = 0.5$ GeV (curve 1)
and $E_m = 0.05$ GeV (curve 2). The model spectrum 
for the case of a 30 km s$^{-1}$ shock in a clump   
with number density  10$^{4}$ cm$^{-3}$ is shown by curve 3 
(see the text).
Interstellar absorption of the radio emission
is as in Fig. 1. 
The internal absorption at the shocked clump is shown for
 EM $\approx$ 110 cm$^{-6}$ pc (solid line radio spectrum) and 
 EM $\approx$ 900 cm$^{-6}$ pc (dashed line radio spectrum).
The dot-dashed curves are the shell emission spectra from Fig. 1 given 
for comparison.} 
\vspace*{0.5cm}

We presented above the results for nonthermal emission of a
strong radiative shock of $v_{S7} \sim 1$ propagating into a
molecular clump.
For the case of  radiative shock interaction with the molecular clumps
in  IC 443, lower shock velocities,  $v_{S7} \sim$ 0.3, are more realistic
(e.g., Cesarsky \etal 1999; 
Chevalier 1999).
The preshock ionization may be dominated mostly by
nonthermal particles in such a shock because UV radiation is inefficient.
A self-consistent model of MHD collisionless shock propagation
into a dense clump requires simultaneous simulations of the MHD turbulence
spectral properties and the nonthermal particle kinetics. Instead, we
suppose that a {\it collisionless} shock transition exists for
$v_{S7} \sim$ 0.3 and $n = 10^4 \cmc$ 
and assume the same diffusion model as described in 
\S 2.  We calculated the upstream gas ionization, finding that the 
ionized fraction
$x \gsim$ 0.9 may hold up to depths about 10$^{15}$ cm$^{-2}$ 
due to shock accelerated electrons. 
The Mach number of the shock with
$v_{S7} \sim$ 0.3 is typically below 5 because of 
heating of upstream gas up to 10$^4$ K by 
accelerated particles and the compression ratio
is lower than 4.  
The maximum energy of accelerated electrons in that case is typically 
below an MeV. 
We illustrate the calculated $\nu F_{\nu}$ spectrum of nonthermal emission 
from a 30 $\kms$ velocity shock ($\delta_t$ = 3 ) interacting with a
molecular clump of density 10$^4 \cmc$ 
by curve 3 in  Fig. 3; this model corresponds to 
$k_{III0} \sim  7 \times$ 10$^{16}$ cm$^2$ s$^{-1}$ and  $E(p_{\ast}$) = 10 keV. 
Such collisionless shocks, if they exist, would provide a
softer spectrum of nonthermal radiation than that for high velocity shocks
with $v_{S7} \gsim$ 1.   
A more comprehensive study is required to
model the nonthermal emission from the low velocity shocks in detail.
A search for  hard X-ray emission correlated with  
molecular emission predicted by the simplified model described above is possible with 
high resolution instruments like {\it Chandra} and {\it XMM}.

\subsection{Hard X-ray Emission}

IC 443 was a target of  X-ray observations with
{ \it HEAO 1} (Petre \etal 1988), {\it Ginga} (Wang \etal 1992),
{\it ROSAT} (Asaoka \& Aschenbach 1994)
and {\it ASCA} (Keohane \etal 1997).
{\it ASCA GIS} observations discovered the localized character of the hard  X-ray
emission (Keohane \etal 1997). Most of the 2-10 keV {\it  GIS} photons
came from an isolated emitting feature and from the southeast
elongated ridge of hard emission. The ridge is also coincident
with the 95\% confidence error circle of the {\it EGRET CGRO} source
(Esposito \etal 1996), while the isolated feature is  outside
the circle (Keohane \etal 1997).
The integrated 2-10 keV flux observed by {\it ASCA GIS} was
(5$\pm 1) \times$ 10$^{-11}$ erg cm$^{-2}$ s$^{-1}$,
representing more than 90\% of the total flux from the SNR
(Keohane \etal 1997). The hard X-ray ridge, as defined by Keohane \etal (1997),
is only a part of the more extended radio shell
discussed in the previous section.
The flux density at 7 keV from the isolated
emitting feature was about
4$\times$ 10$^{-5}$ photons  cm$^{-2}$ s$^{-1}$ keV$^{-1}$  and that
from the ridge was 
2$\times$ 10$^{-5}$ photons  cm$^{-2}$ s$^{-1}$ keV$^{-1}$.
{\it HEAO A-2} measured a flux at 2--10 keV of
(7$\pm 1) \times$ 10$^{-11}$ erg cm$^{-2}$ s$^{-1}$
(Petre \etal 1988), while {\it Ginga} measured a 2--20 keV
flux of about 9$\times$ 10$^{-11}$ erg cm$^{-2}$ s$^{-1}$
(Wang \etal 1992).
These numbers are consistent with the assumption that the integrated
flux is dominated by the extended soft component coming from the central
parts of the remnant.

The soft X-ray 0.2--3.1 keV surface brightness map of IC 443 from the
{\it Einstein} Observatory (Petre \etal  1988)
shows bright features in the northeastern part of the remnant.
 The presence of nearly uniform X-ray emission from the central part
of the remnant is a characteristic feature of mixed-morphology 
SNRs (Rho \& Petre 1998). It might be due to the effect of
thermal conduction  (Cox \etal 1999; Shelton \etal 1999),
although Harrus \etal (1997) argued for an alternative radiative shock model
for W44, another
mixed-morphology SNR. That feature in our model corresponds
to the emission from hot (T$\lsim 10^7$ K), low density gas interior
to the shock (region {\it V}).

In the radiative shock model with direct injection
from the thermal pool 
and second order Fermi acceleration
in postshock cooling layer,
most of the hard X-ray emission ($\gsim 7$ keV)
comes from the shell - radiative shock structure with a rising
$\nu F_{\nu}$ X-ray spectrum (curve 1 in Fig. 1 for the  shell
and curves 1 and 2 in Fig. 3 for a shocked molecular clump).
In addition to the shell, there is
a  more extended source of hard X-rays
from inverse Compton emission.
It is the dominant
source of hard X-rays in the scenario with cosmic ray electron
reacceleration (Fig. 2) and in the case of a lack of second order
Fermi acceleration in the radiative shell (see curve 2 in Fig. 1).
In the scenario with reacceleration of cosmic ray
electrons,
it is impossible to obtain the 7 keV flux observed by {\it ASCA}
from the IC 443 ridge.
Thus, a resolved hard X-ray image of the IC 443 remnant may be very
informative in determining the electron acceleration scenario.

The isolated emitting feature in the southern part of IC 443
detected by {\it ASCA} 
and brightest around 7 keV, 
has a flat spectrum low frequency radio continuum source (Green 1986)
and is close to a region of excess H$_2$ emission.
High spatial resolution observations
of this isolated emitting feature are required to
decide between a low-luminosity pulsar nebula and shock interaction
models (e.g., radiative shock interaction with a dense molecular
clump, Fig. 3). Observations of the hard emission
with the sensitivity and resolution
available with  {\it BeppoSAX, Chandra XRO, XMM} and {\it GLAST}
would  be required to constrain the extension and spectral
properties of the hard X-ray emission from IC 443.

\subsection{Gas Ionization by Nonthermal Electrons}

Nonthermal particles accelerated by a MHD collisionless shock wave  
provide an efficient ionizing agent. We calculated 
above the nonthermal emission generated by accelerated energetic electrons; 
one can also calculate the  ionization produced by the same 
electrons in the molecular cloud. This might connect the high energy 
 observations  with radio, IR and optical 
emission from shocked atomic and molecular gas. 
To estimate the gas ionization due to accelerated electrons 
in the radiative shock structure we used the electron spectra calculated with the
kinetic model of electron acceleration and propagation 
described in \S 2 and electron ionization cross sections 
from  Appendix B. 
In Fig. 4, we present the calculated ionization rate $\zeta_e$ 
and estimated ionized fraction $x$ 
 in the shell 
of a radiative shock due to accelerated electrons for the 
case of $v_{S7} = 1.5$ for both models of large-scale turbulence in
the postshock cooling layer considered in \S 3.1.

\vspace*{-1.2cm}
\centerline{\null}
\vskip4.55truein
\includegraphics{fig4.pstex}
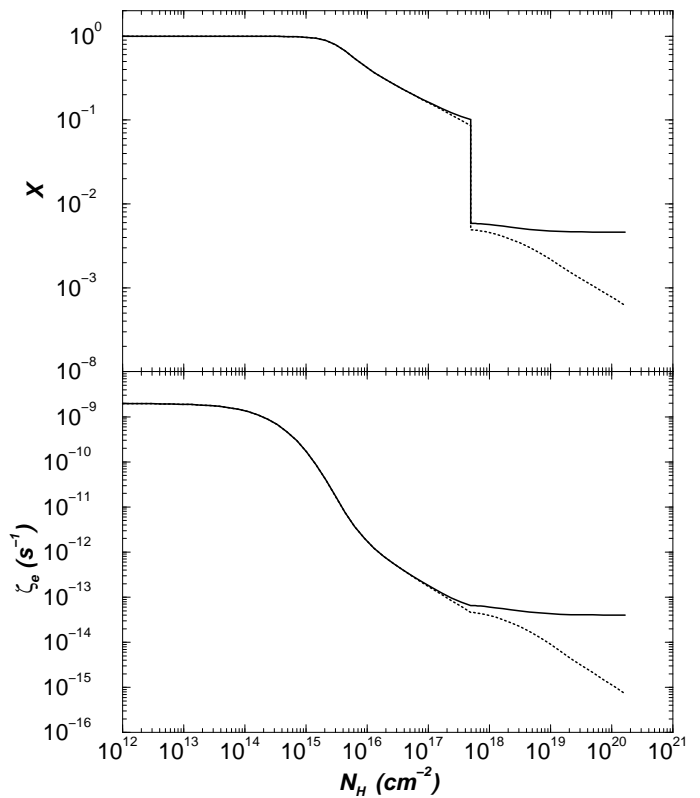
\figcaption{Postshock ionization rate 
due to nonthermal electrons $\zeta_e$ (lower panel)
and the estimated ionization fraction $x$ (upper panel)
versus total H column density.   The solid lines  correspond to
the model with $E_C$ = 120 keV (see curve 1 in Fig. 1)
and the dotted lines
to $E_C$ = 2 GeV (see curve 2 in Fig. 1).}
\vspace*{0.5cm}
\label{fig:fig4}

There should also be
 contributions to $x$ from accelerated nucleons and from UV radiation which 
are not included in Fig. 4. We accounted for only radiative 
recombination, assuming a small molecular fraction in the 
radiative shell at  column densities below 2 $\times 10^{20}$ cm$^{-2}$. 

The solid curves 
in both panels of Fig. 4 correspond
to the case of fully developed
Alfvenic turbulence in the postshock cooling layer,
which provides efficient  second order Fermi acceleration
making Coulomb losses relatively unimportant above $E_C$ = 120 keV.
This case corresponds to  curve 1 in Fig. 1 for  emission
from the shell with substantial hard X-ray emission.
The dotted lines in Fig. 4 correspond to the case of a lack of large scale
MHD turbulence where the Coulomb losses overcome the second order
Fermi acceleration up to $E_C$ = 2 GeV. The nonthermal emission
expected in this case is illustrated by curve 2 in Fig. 1.
The radiative shell ionization by nonthermal
electrons is sensitive to the turbulent structure of the postshock
cooling layer and there is a correlation between the hard X-ray spectrum
and the ionization structure of the radiative shell.
We shall consider the effect of nonthermal particles
on the ionization structure of a shocked clump elsewhere.

\subsection{Energy in Nonthermal Electrons}

The energy of the electron component for the parameter set used to
compute the spectra in Fig. 1 is substantial.  The power required for
electron acceleration and maintenance for curve 1 in Fig. 1, $\sim 1.5
\times 10^{37}$ erg s$^{-1}$, is high because of the strong Coulomb
losses of keV electrons in the dense ionized medium, but it is
consistent with the estimate of the total radiative losses from IC
443, dominated by infrared emission (Mufson \etal 1986; Burton \etal
1990).  A similar power is required for both models 1 and 2 in Fig. 3,
where radiative shock - molecular clump interaction is illustrated.
For curve 2, the power requirements are less, $\sim 6 \times 10^{36}$
erg s$^{-1}$.  The implied efficiency of power conversion from the MHD
shock flow (which can be estimated as 3 $\times 10^{38}$ erg s$^{-1}$)
to the nonthermal electrons is about 5\% in these cases.  The model
with $v_{S7} \approx 1$ would provide emission similar to that shown
in Fig. 1, but requires about three times higher efficiency.  This is
higher than the electron acceleration efficiency estimated from GeV
regime cosmic rays observed near the Earth, but evolved SNRs
interacting with dense molecular gas probably cannot be considered as
the main source of the observed Galactic cosmic rays because of the
relatively low maximum energies of accelerated particles.

The pressure of the nonthermal
electron component downstream of the shock (region {\it IV})
is ${\cal P}_e \approx$ 7$\times$ 10$^{-11}$ erg cm$^{-3}$.
The magnetic pressure in the dense shell is about
1.5 $\times$ 10$^{-10}$ erg cm$^{-3}$, which is higher than the
thermal gas pressure.
The uniform ($\sim$ 60 $\mu$G) and stochastic magnetic field
components  dominate the total pressure in the shell.

 Let us consider the energy requirements for the scenario with
 electron injection from preexisting cosmic rays (Fig. 2).
 The nonthermal electron pressure in this case is
 ${\cal P}_e \approx$ 5.2$\times$ 10$^{-11}$ erg cm$^{-3}$ (for $E_{crm}\leq$ 8
 MeV), similar to 
 that for the model with  injection from the thermal pool, because it 
 is dominated by GeV  particles. An  advantage of 
 this scenario is that much less power is required. 
 It is  about 4 $\times$ 10$^{35}$
 erg s$^{-1}$, which is less than 10\% of that for the scenario with 
  injection from the thermal pool. The difference is because Coulomb  
losses are unimportant for the high energy electrons involved
in that scenario.
However, a potential problem is the relatively low $E_{crm}$
required. The lifetime of a 20 MeV electron against ionization 
losses in a neutral medium of number density $\sim$ 1 cm$^{-3}$ is about 
10$^6$ years. This implies that the source of MeV cosmic ray electrons
should exist within about 100 pc of the molecular cloud if  
the diffusion coefficient is $\sim 3\times$ 10$^{26}$ cm$^2$ s$^{-1}$ at 
these energies.

\section{\bf DISCUSSION AND FUTURE PROSPECTS}
 
For individual  supernova
remnants, the ambient density is an important parameter
that can be estimated from multiwavelength observations 
(Chevalier 1999). Another important parameter for modeling 
the nonthermal particles in a SNR in a molecular cloud is  
the collisionless MHD turbulence spectrum, particularly in the 
radiative shock cooling layer.   A substantial level of MHD 
collisionless turbulence could overcome Coulomb losses 
in the dense plasma downstream from the radiative shock. 
An accurate model of MHD turbulence in the postshock cooling layer 
is not available now. Thus we considered both limiting cases described in 
the \S~2 and show in  Fig. 1 the uncertainty introduced by the
lack of data concerning  the MHD turbulence properties.  
Coulomb losses are important in the postshock cooling region 
only for electrons with energies
below $E_C \approx$ 20 keV if the large scale turbulent velocity has a
substantial longitudinal component of $\sim 20 \kms$.  
A higher value ($E_C \gsim$ 8 MeV) is expected in the absence of a 
longitudinal component 
of large scale turbulence in the postshock cooling layer because
the Coulomb losses are overcome by resonant interaction with Alfven 
waves. Since the Alfven velocity is relatively low  ($\sim 5\kms$), 
the value of $E_C$ is much higher in that case.
We used a conservative minimum value of $E_C$ = 120 keV for the radiative 
shock structure described above, although one could expect 
$E_C \sim$ 20 keV in the most favorable case. The nonthermal emission of 
SNR in a molecular cloud (especially in the hard X-ray regime) 
and ionization structure of the radiative shock 
are sensitive to the MHD turbulence model.   Spatially resolved 
observations of the nonthermal emission from  SNRs may be able to
 constrain models of MHD turbulence. 
 
All of the SNR candidates from the {\it EGRET} list:
IC 443,  $\gamma$ Cyg, W44, and
Monoceros (Esposito \etal 1996) are old remnants
interacting with molecular clouds.
The \gr emission from electrons accelerated by the
radiative shock calculated for the IC 443 parameters
presented in Figs. 1 and 2 is in good agreement with that
observed by {\it EGRET} (Esposito \etal 1996), 
as well as with the upper limits established 
 by {\it Whipple} (Buckley \etal 1998).
The high energy $\gamma$-rays ($\gsim 50$ MeV) from the radiative shell should be
spatially correlated
with the radio emission.
The {\it EGRET} telescope detected an extended excess
(95\% confidence circle $\sim 42^{\prime}$ in radius)
from the Monoceros SNR correlated with the 1.42 GHz
radio emission (Esposito \etal 1996). This is in accordance with
the model of a leptonic origin of GeV  emission from extended
SNRs interacting with clouds (de Jager \& Mastichiadis 1997).
{\it GLAST}
(Gamma-ray Large Area Space Telescope) will be an excellent
instrument for future $\gamma$-ray observations,
because it will have the capability of spatially resolving
the $\gamma$-ray emission from SNRs.

Spatially resolved observations of
W44, IC 443, 3C391 and some other mixed-morphology SNRs from the list
given by Rho \& Petre (1998) with {\it BeppoSAX, Chandra XRO, XMM}
would be very valuable tools to test our model.
The hard X-ray detector (HXD) aboard the
forthcoming {\it ASTRO-E} mission could be used for observations of the
hard 10--700 keV continuum with a field of view of
0.8$^{\circ}$ FWHM
at 60 keV and 2.8$^{\circ}$ at 500 keV from the extended hard X-ray
 emission from IC 443, W44 and the Monoceros SNR predicted by
the radiative shock model (see Fig. 1 and Fig. 3).   Spatially
resolved spectra are needed to distinguish the 
 shell emission  from the shocked clump emission.

Molecular clumps interacting with moderately fast radiative
shocks are also expected
to be  sources of hard X-rays and MeV  $\gamma$-rays (up to 100s of MeV).  
Nonthermal continuum radio emission (100s of MHz) with 
a sharply rising $\nu F_{\nu}$ spectrum (Fig. 3) 
and a time dependent cut-off frequency is expected in the fast shock model. 
 Due to potentially
substantial internal free-free absorption, the spectrum of radio
emission from a localized clump might constrain the ionized gas density
in the clump. The MeV \gr spectrum
of a localized clump may be resolved with  the
forthcoming \gr missions {\it INTEGRAL}
(e.g., Sch\"onfelder 1999; Winkler 1999) and {\it GLAST}.
With an expected
angular resolution about 12$^{\prime}$ FWHM, imager {\it IBIS} aboard
{\it INTEGRAL} would allow detection of hard X- ray ($\gsim$ 50 keV)
emission from  molecular clouds within $\sim$ 1 kpc.
A comparison  of hard emission spectra with the radio spectrum can
provide  valuable information about the density and magnetic
field in a clump. A possible 
variability on a
timescale of a few years for clump hard emission and radio emission
(on a longer time scale) could further constrain the model.

For low velocity (below $\sim 30 \kms$)  shocks interacting with a dense 
($\sim 10^4 \cmc$),  magnetized ($B \lsim$ 0.5 mG) molecular clump,
one may expect hard X-ray emission below the MeV regime correlated with 
the regions of  molecular emission of shocked gas. Nonthermal 
radio continuum is not expected in the case of low velocity shocks 
because of a lack of accelerated relativistic electrons at
sufficiently high energies. 

Finally, we note that the high density of energetic particles in the
vicinity of a shock wave in a molecular cloud can affect the ionization
and thermal properties of the gas.
Rich molecular spectra have been observed from IC 443 and
3C391 (Reach \& Rho 1999).
It will be interesting to see whether there is a signature of the
presence of nonthermal particles that can be discerned from
the molecular spectra.

\acknowledgments
A.M.B  thanks Hans Bloemen for useful discussions. We are grateful to
M.G. Baring for comments and for
making available the corrections to the
e-e bremsstrahlung formulae and to the referee for very constructive comments.
The work of A.M.B and Yu.A.U was supported by the INTAS grant 96-0390
and that of R.A.C. by NASA grant NAG5-8232.

\appendix

\section{\bf Appendix A: Radiative Processes}

Given the electron distribution function, the high energy emission 
flux $J(E_{\gamma})$ at a distance $R$ from the source 
can be calculated from the equation:
\begin{equation}
J(E_{\gamma})=\frac1{4\pi R^2}\int\limits_V dV \frac{dn_{\gamma}(E_{\gamma},
\overrightarrow{r})}{dt}
\end{equation}
where  
${dn_{\gamma}(E_{\gamma},\overrightarrow{r})}/{dt}$ is the emissivity,
which here includes bremsstrahlung,
synchrotron radiation and the inverse Compton effect.
We generally assumed a pure hydrogen composition in our calculations.

High energy particles penetrating through a partially ionized medium produce
photons due to interactions with atoms, electrons and ions. 
We assumed a Maxwellian distribution for the ambient matter and in most cases 
took the target ambient particles to be at rest. 
Then, the emissivity can be calculated from
\begin{equation}
\frac{dn_{\gamma}(E_{\gamma})}{dt}=4\pi \left( 
\int\limits_{E_{\gamma}}^{\infty} dE_e
N_p \sigma_{ep}(E_{\gamma},E_e) J(E_e)  +
\int\limits_{E_{min}}^{\infty} dE_e
N_e \sigma_{ee}(E_{\gamma},E_e) J(E_e) \right),
\end{equation}  
where $J(E_e)$ is the electron flux measured in units
s$^{-1}$ cm$^{-2}$ keV$^{-1}$ sr$^{-1}$ ,
 $\sigma_{ep}$ and $\sigma_{ee}$ are differential cross sections
of one photon emission due to electron-proton and electron-electron
interactions integrated over angles, and 
$E_{min}$ is the minimum energy of
incoming electron that could generate a photon with energy $E_{\gamma}$. 
For e-p interactions, $E_{min}= E_{\gamma}$ is a  good approximation.
The correction due to finite proton mass is less than the uncertainty
in the cross section.
In the case of e-e interactions, we used the following equation for $E_{min}$: 
\[
E_{\gamma}=E_{min} m_e c^2 / \left[2 m_e c^2 + E_{min} - 
\sqrt{E_{min} (E_{min} + 2m_e c^2)}\right].
\]
The inverse bremsstrahlung contribution is negligible unless there is a very high 
(far in excess of 100) ratio of protons to electrons.

We used the following cross section approximations in our calculations:

\subsection{\bf e-e bremsstrahlung.}

Haug (1975) has obtained a general formula for the e-e bremsstrahlung 
cross section,
but it is more convenient to use simplified approximations.
We used relativistic and non-relativistic
asymptotic expressions, which were matched in the intermediate energy region 
(see Baring \etal 1999).

In the relativistic limit, we used approximate formulae given by Baier \etal
(1967) with a correction factor from Baring \etal 
(1999):\footnote{Equations (A4) and (A6) are somewhat different from that given by
Baring \etal (1999).
M.G. Baring (private communication) made available the corrections to their
original formulae.}
\begin{equation}
\sigma_{ee}(E_e,E_{\gamma}) = (\sigma_1+\sigma_2) A(\epsilon_{\gamma},\gamma_e),
\end{equation}
where $\gamma_e=(E_e+m_e c^2)/m_e c^2$, $\epsilon_{\gamma}=E_{\gamma}/m_e c^2$,
\begin{equation}
\sigma_1(\gamma_e,\epsilon_{\gamma}) = 
\frac{4 r_o^2 \alpha}{\epsilon_{\gamma}}\left[1+\left(\frac13-
\frac{\epsilon_{\gamma}}{\gamma_e - 1}\right)
\left(1-\frac{\epsilon_{\gamma}}{\gamma_e - 1}\right)\right]\left[\ln\left(2(\gamma_e - 1)
\frac{\gamma_e- 1- \epsilon_{\gamma}}{\epsilon_{\gamma}}\right)-\frac12\right],
\end{equation}
\begin{equation}
\sigma_2=\frac{r_o^2 \alpha}{3 \epsilon_{\gamma}}\left\{
\begin{array}{lr}
\left[16 (1-\epsilon_{\gamma}+\epsilon_{\gamma}^2)\ln\left(\frac{\gamma_e}
{\epsilon_{\gamma}}\right) -\frac1{\epsilon_{\gamma}^2}+
\frac3{\epsilon_{\gamma}}-4+4\epsilon_{\gamma}-8\epsilon_{\gamma}^2\right.  & \\
\left.-2(1-2\epsilon_{\gamma})\ln(1-2\epsilon_{\gamma})\left(
\frac1{4\epsilon_{\gamma}^3}-\frac1{2\epsilon_{\gamma}^2}
+\frac3{\epsilon_{\gamma}}-2+4\epsilon_{\gamma}\right)\right], &  
\epsilon_{\gamma}\le\frac12,\\
\frac2{\epsilon_{\gamma}}\left[\left(4-\frac1{\epsilon_{\gamma}}
+\frac1{4\epsilon_{\gamma}^2}\right)\ln(2\gamma_e)-2 
+\frac2{\epsilon_{\gamma}}-\frac5{8\epsilon_{\gamma}^2}\right], &
\epsilon_{\gamma}>\frac12,
\end{array} \right.
\end{equation}
$r_o={e^2}/{m_e c^2}$ is the classical electron radius,
$\alpha = {e^2}/{\hbar c}$, and
\begin{equation}
A(\epsilon_{\gamma},\gamma_e)=1-\frac{10}{3} \frac{(\gamma_e-1)^{\frac15}}
{\gamma_e+1} \left(\frac{\epsilon_{\gamma}}{\gamma_e}\right)^{\frac13}.
\end{equation}
According to Baring \etal  (1999),
these formulae agree with those from Haug (1975) with an accuracy $\sim 10 \%$ for
 electrons of energy $\ga 5$~MeV .

In the non-relativistic limit, we used equations from Garibyan (1952) and 
Fedyushin (1952)  with correction coefficients from Baring \etal  (1999):
\begin{equation} 
\sigma_{ee}=\frac{4 r_o^2 \alpha}{15\epsilon_{\gamma}}
F\left(\frac{4 \epsilon_{\gamma}}{\gamma_e^2-1}\right)
\end{equation}
where the photon energy  is  in the range 
$0<\epsilon_{\gamma}<\frac14(\gamma_e^2-1)$
and  $F(x)$ ($0<x<1$) has the form:
\begin{equation}
\begin{array}{lll}
F(x) & = & B(\gamma_e)\left[17-\frac{3 x^2}{(2-x)^2} - C(\gamma_e,x)\right]
\sqrt{1-x}\\
 & & +\left[12(2-x)-\frac{7x^2}{2-x}-\frac{3x^4}{(2-x)^3}\right]
 \ln\left(\frac{1+\sqrt{1-x}}{\sqrt{x}}\right).
\end {array} 
\end{equation} 
The correction coefficients are:
\begin{equation}
B(\gamma_e)=1+\frac12(\gamma_e^2-1)\ ;\ \ \ \ C(\gamma_e,x)=
\frac{10x\gamma_e\beta_e(2+\gamma_e\beta_e)}{1+x^2(\gamma_e^2-1)}
\end{equation}
Baring \etal  (1999) found less than a
$10 \%$
difference between equation
 (A7)  and the expressions of Haug (1975) for electron energies
$E_e \la 500$ keV.

\subsection{\bf e-p bremsstrahlung.}

In the
non-relativistic limit, we took the e-p bremsstrahlung cross section
from Akhiezer \& Berestetsky (1957):
\begin{equation}
\sigma_{ep}=\frac{16 r_0^2 Z^2 \alpha m_e^2 c^2}{3 p_1^2} \frac{4\pi^2\zeta_1\zeta_2}
{(e^{2\pi\zeta_1}-1)(1-e^{-2\pi\zeta_2})}\frac1{\epsilon_{\gamma}}
\ln\left(\frac{p_1+p_2}{p_1-p_2}\right),
\label{eq:BornAh} 
\end{equation} 
which is valid if $\zeta_1 \ll 1$; here $\zeta_i=\alpha Z / \beta_i$.

For the relativistic limit, we used Born approximation formulae from 
Akhiezer \& Berestetsky (1957):
\begin{equation}
\begin{array}{ll}
\sigma_{ep}= & r_o^2 Z^2 \alpha \frac{p_2}{p_1} \left\{\frac43-
2\epsilon_1\epsilon_2\frac{p_1^2+p_2^2}{p_1^2 p_2^2} +
m_e\left(\frac{\eta_1\epsilon_2}{p_1^3} + \frac{\eta_2\epsilon_1}{p_2^3}-
\frac{\eta_1\eta_2}{p_1 p_2}\right) + \right.\\
  & \left. L\left[\frac83 
\frac{\epsilon_1 \epsilon_2}{p_1 p_2} + \frac{\epsilon_{\gamma}^2}{p_1^3 p_2^3}
(\epsilon_1^2 \epsilon_2^2+p_1^2 p_2^2)+ \frac{m_e^2 \epsilon_{\gamma}}{2p_1p_2}
\left(\eta_1\frac{\epsilon_1 \epsilon_2 + p_1^2}{p_1^3}-\eta_2
\frac{\epsilon_1 \epsilon_2+p_2^2}{p_2^3} + 
\frac{2\epsilon_{\gamma}\epsilon_1\epsilon_2}{p_1^2 p_2^2}\right)\right]\right\},
\end{array}
\label{eq:LanAh}
\end{equation}
where $L=\ln\left(\frac{p_1^2+p_1 p_2 - \epsilon_1 \epsilon_{\gamma}}
{p_1^2-p_1 p_2 - \epsilon_1 \epsilon_{\gamma}}\right)=2\ln\left(
\frac{p_1^2+p_1 p_2 - m_e^2}{m_e \epsilon_{\gamma}}
\right)$,
$\eta_1=\ln\left(\frac{\epsilon_1+p_1}{\epsilon_1-p_1}\right)=
2\ln\left(\frac{\epsilon_1+p_1}{m_e}\right)$, and $\eta_2=2\ln\left(
\frac{\epsilon_2+p_2}{m_e}\right)$.
In the velocity range $Z\alpha(\beta_2^{-1}-\beta_1^{-1})$, we included 
the Elwert factor from Koch \& Motz (1959):
$f_E=\frac{\beta_1 (1-exp[-2\pi Z \alpha / \beta_1)])}
{\beta_2 (1-exp[-2 \pi Z \alpha / \beta_2)])}$. 

In  equation (A11), $ c = 1$ is assumed.
In the intermediate energy range, we matched 
equation (A10) and equation (A11). 

\subsection{\bf Synchrotron radiation.}

We used standard formulae for synchrotron radiation  
in our calculations (e.g., Ginzburg 1979). 
The emissivity can be written as:
\begin{equation}
\begin{array}{lll}
E_{\gamma} \cdot \frac{dn_{\gamma}(E_{\gamma})}{dt} & = &
\frac{\sqrt{3}\ \alpha e}{ m_e c}
\int d\chi dE N(E,\chi) H \sin^2(\chi) F\left(
\frac{\nu}{\nu_c}\right),\\
F(x) & = & x\int\limits_x^{\infty}K_{5/3}(\eta) d\eta,
\end{array}
\label{eq:synhtot}
\end{equation} 
where $\chi$ is the angle between the electron velocity
and the magnetic field,
$N(E,\chi)$ is the electron distribution function over energy and angle,  
$\nu_c=\frac{3eH\sin(\chi)}{4\pi m_e c}
\left(\frac{E}{m_e c^2}\right)^2$, and $K_{5/3}(x)$ is the 
McDonald function.
Integrating equation (12) over angle under the assumption of a chaotic
magnetic field orientation, we obtain the equation
\begin{equation}
E_{\gamma} \cdot \frac{dn_{\gamma}(E_{\gamma},
\overrightarrow{r})}{dt} =
\frac{\sqrt{3}\ \alpha e}{2 \pi m_e c}
\int dE N(E,\overrightarrow{r}) H_{\bot} F\left(
\frac{\nu}{\nu_c}\right),
\end{equation} 
where $N(E,\overrightarrow{r})$ is the electron distribution function as
a function of energy and position and
$H_{\bot}=\sqrt{2/3} H(\overrightarrow{r})$ is the averaged
magnetic field.

\subsection{\bf Inverse Compton emission}

High energy electrons colliding with  photons result in high energy 
photon production (e.g., Jones 1968; Gaisser \etal  1998;
Sturner \etal  1997; Baring \etal  1999).
If the electron energy $E_e \ll \frac{m_e^2 c^4}
{4 E_{\gamma}}\approx \frac{ 6\cdot 10^{10} }
{E_{\gamma} (\mbox{eV})}~$eV ($E_{\gamma}$ is the energy of the emitted photon), 
then the formulae in the Thompson limit are (e.g., Ginzburg 1979):
\begin{equation}
\frac{dn_{\gamma}(E_{\gamma})}{dt}=\sqrt3 \pi \sigma_T m_e c^2 \int dE_{ph}
\sqrt{\frac{E_{\gamma}}{E_{ph}}}
{N_{ph}(E_{ph},\overrightarrow{R})} J_e\left(
m_e c^2 \sqrt{\frac{3E_{\gamma}}{4 E_{ph}}},\overrightarrow{R}\right), 
\end{equation}
where $\sigma_T=\frac{8\pi}3 \left(\frac{e^2}{m_e c^2}\right)^2$ is the 
Thompson cross section, $N_{ph}(E_{ph})$ is the background photon energy spectrum, and
$J_e(E)$ is the incoming electron flux.

If the electron energy  
$E \ge \frac{m_e^2 c^4}{4 E_{\gamma}}\approx \frac{ 6\cdot 10^{10} }
{E_{\gamma} (\mbox{eV})}~$eV,  we must use the Klein-Nishina 
formulae (integrated over angles): 
\begin{equation}
\frac{dn_{\gamma}(E_{\gamma})}{dt}=4\pi  
\int\nolimits  dE_e dE_{ph}
N_{ph}(E_{ph}) \sigma_{KN}(E_{\gamma},E_e,E_{ph}) J(E_e), 
\end{equation}  
\begin{equation}
\sigma_{KN}(E_{\gamma},E_e,E_{ph})=
\frac{2\pi r_o^2}{\epsilon_{ph} \gamma_e^2} \left[2q\ln(q) +1+q -2q^2+
\frac{q^2 (1-q) \Gamma^2}{2(1+ q\Gamma)}\right],
\end{equation}
\[
\mbox{ where } \Gamma= 4\epsilon_{ph}\gamma_e,~
 q=\frac{\epsilon_{\gamma}}{(\gamma_e-\epsilon_{\gamma}) \Gamma},
\mbox{      } 0\le q\le 1.
\]
In the above, $\epsilon_{\gamma}=E_{\gamma}/m_e c^2$, 
$\epsilon_{ph}=E_{ph}/m_e c^2$,
$\gamma_e$ is the electron Lorentz factor, 
and $r_o=e^2/m_e c^2$ is the classical electron radius. 
The interstellar background photon spectrum, $N_{ph}(E_{ph},R)$, was adopted from  
Mezger \etal  (1982) and  Mathis \etal  (1983) 
taking into account infrared data
from Saken \etal  (1992) for the case of IC443  (see also Gaisser \etal  1998).

\subsection{\bf Free-free absorption}

Radio waves propagating in the partially ionized thermal plasma are
 subject to thermal free-free absorption. For $h\nu \ll kT$
(the Rayleigh -- Jeans regime), the free-free absorption coefficient
$\alpha_{\nu}^{ff}$ (cm$^{-1}$) can be calculated from
\begin{equation}
\alpha_{\nu}^{ff} = \frac{4e^6}{3m_ekc} \left(\frac{2 \pi}{3km_e}\right)^{1/2}
T^{-3/2} Z^2 n_e n_i \nu^{-2} \bar{ g_{ff}}.
\end{equation}
Using an approximate formula for the Gaunt factor $\bar{ g}_{ff}(\nu,T)$
from Rybicki \& Lightman (1979) appropriate to our parameter range, we
obtained a simplified expression for the optical depth $\tau(\nu, T)$
\begin{equation}
\tau \approx 0.022~ \nu_{100}^{- 2.1} T_{100}^{-1.34} EM ,
\end{equation}
where  the radio wave frequency $\nu_{100}$ is measured in units of 100 MHz,
the plasma temperature in units of 100 K is $T_{100}$, and the emission measure
$EM$ is in  units of cm$^{-6}$ pc.

\section{\bf Appendix B: Electron Ionization }

The primary ionization rate $\zeta_e$ due to electron impact was calculated
from the equation:
\begin{equation}
\zeta_e =4\pi \int\limits_{E_{min}}^{\infty} \!dE\,
\sigma_{eion}(E) J_e(E),
\end{equation}  
where $J_e(E)$ is the electron flux measured in units
s$^{-1}$ cm$^{-2}$ keV$^{-1}$ sr$^{-1}$ ,
 $\sigma_{eion}$  is the differential cross section
for hydrogen atom primary ionization by electron-atom
interactions integrated over atomic electron states, and
$E_{min}$ is the minimum energy of the
ionizing incoming electron.
 
In the 
regime below 1 keV, we obtained the following  fit to  
the ionization cross section $\sigma_{eion}$ using experimental data 
from Fite \& Brackmann (1958): 
\begin{equation}
\sigma_{eion}(E)=\pi a_0^2
A_1\, \exp(-bA_4-b^3A_5-b^5A_6)/{E({\rm eV})^{A_3}},
\end{equation}
where $b={A_2}/{E({\rm eV})}$ and $E$(eV) is the incoming electron energy 
measured in eV. 
The fitting coefficients  are $A_1$=76.11, $A_2$=14.34, $A_3$=0.89, $A_4$=3.82, 
        $A_5$=--2.55, and $A_6$=4.5. $a_0={\hbar}^2/m_e e^2$ is the
Bohr radius.

At high energies  ($\gsim 1$ keV), the Born  approximation provides  good 
accuracy (e.g., Mott \& Massey 1965): 
\begin{equation}
\sigma_{eion}=\int^{\kappa_{max}}_0\int^{K_{max}}_{K_{min}}I_{0\kappa}(K)\,dK
\,d\kappa,
\end{equation}
where
\begin{eqnarray}
   I_{0\kappa}(K)
   & = & \frac{  2 \pi\, 2^{10}\, \kappa  }{
                \rule{0em}{3.2ex}a_0^2 \,k^2 \, K  } \;
         \frac{\mu^6
                \left[ \rule{0em}{2.5ex} K^2 + ( \mu^2+\kappa^2 )/3 \right]
                  }{ \left[ \mu^4+2\mu^2
                         \left( \rule{0em}{2ex} K^2+\kappa^2 \right)
                       + \left( \rule{0em}{2ex} K^2-\kappa^2\right)^2
                     \right]^3
               } \nonumber \\
   & \times & \exp \! \left[
        -\frac{2\mu}{\kappa}\; {\rm arctg}\!
        \left(  \frac{2\mu\kappa}{K^2-\kappa^2+\mu^2} \right)
                    \right]
        \frac1{1-\exp \! \left( -2\pi \mu / \kappa \right) },
\end{eqnarray}
$\mu=Z/a_0$, $k(E)$ is the incoming electron wave number, $k_{\kappa}$ is the
outgoing electron wave number, and $\kappa$ is the emitted atomic 
electron wave number. 
In addition, $K=|\overrightarrow k - \overrightarrow k_{\kappa}|$,
$K_{max}=k+k_{\kappa}$, and $K_{min}=k-k_{\kappa}$.

We found a convenient approximation providing a reasonably accurate 
fit (better than 10\%  above 1 keV and better than 5\% above 3 keV) 
to  equations (B3) and (B4):
\begin{equation}
\sigma_{eion}(E)=\pi a_0^2~ P_1~ E(eV)^{-P_2} ,
\end{equation}
where $P_1$=97.72 and $P_2$=0.93.

The gas ionization fraction $x = n_i/(n_i + n_n)$ (we consider
here a hydrogen plasma)
can be expressed through the ionization rate $\zeta$ and recombination
rate $\alpha_r(T)$ as:
\begin{equation}
  x = \left( \rule{0em}{2.5ex} s^2/4 + s \right)^{0.5} - s/2
\end{equation}
where $s = \zeta/(\alpha_r n)$.
We used the radiative recombination rates given by Spitzer (1978)
and dissociative recombination rates from Schneider \etal (1994).


\clearpage

\end{document}